\documentclass[12pt]{article}
 \usepackage{amssymb}

 \newcommand{\beq}[1]{\begin{equation}\label{#1}}
 \newcommand{\eeq}{\end{equation}}
 \newcommand{\bear}[1]{\begin{eqnarray}\label{#1}}
 \newcommand{\ear}{\end{eqnarray}}

 \catcode`\@=11 \@addtoreset{equation}{section}\catcode`\@=12

 \newcommand{\np}{ {\newpage } }
 \newcommand{\R}{ \mathbb{R} }
 
 \newcommand{\e}{ \mbox{\rm e} }
 \newcommand{\eps}{ \varepsilon }
 \newcommand{\p}{\partial}
 \newcommand{\btd}{\bigtriangledown}
 \newcommand{\btu}{\bigtriangleup}

 \newcommand{\fnm}{\footnotemark}
 \newcommand{\fnt}{\footnotetext}

\begin{document}

 \author{H. Dehnen\\    Konstanz, Germany  }

 \begin{center} \large \bf

 Composite electric $S$-brane  solutions with maximal number of
 branes and several scalar fields

 \end{center}

 \vspace{0.63truecm}

 \bigskip

 \begin{center}

 \normalsize \bf

   H. Dehnen\fnm[1]\fnt[1]{Heinz.Dehnen@uni-konstanz.de},

 \bigskip
 \it Universit$\ddot{a}$t Konstanz, Fakult$\ddot{a}$t f$\ddot{u}$r
 Physik, Fach  M 568, D-78457, Konstanz \\

 \bigskip

 \bf   V.D. Ivashchuk\fnm[2]\fnt[2]{rusgs@phys.msu.ru}
   and
   V.N. Melnikov\fnm[3]\fnt[3]{melnikov@phys.msu.ru},

 \bigskip

 \it  Center for Gravitation and Fundamental Metrology,
  VNIIMS and  Institute of Gravitation and Cosmology,
  Peoples' Friendship University of Russia,
  3-1 M. Ulyanovoy Str., Moscow, 119313, Russia

 \end{center}

  \begin{abstract}

 A $(n+1)$-dimensional cosmological model
 with a set of scalar fields  and  antisymmetric $(p+2)$-form
 is considered. Some of scalar fields may have negative kinetic
 terms, i.e. they may describe ``phantom'' fields.
 For  certain odd dimensions ($D = 4m+1 = 5, 9, 13, ...$) and
 $(p+2)$-forms ($p = 2m-1 = 1, 3, 5, ...$) and  non-exceptional
 dilatonic coupling vector $\vec{\lambda}$ we obtain
 cosmological-type solutions to the field equations.
 These solutions  are characterized by self-dual or anti-self-dual charge
 density forms $Q$ (of rank $2m$) and may describe the maximal set of branes
 (i.e. when all the branes have  non-zero charge densities).  Some properties
 of these solutions are considered, e.g. Kasner-like  behavior,
 the existence of non-singular (e.g. bouncing) solutions and
 those with acceleration. The solutions with bouncing
 and acceleration take place when at least there is one
 ``phantom'' field in the model.

 \end{abstract}

 % \keywords{S-branes, Kasner behavior, power-law expansion,
 %            self-dual charge density form}

 % \preprint{hep-th/0508xxx}

 %\title{Composite electric $S$-brane \\
 %solutions with maximal number of branes and several scalar fields}

 \np

 %%%%%%%%%%%%%%%%%%%%%%%%%%%%%%%%%%%%%%%%%%%%%%%%%%%%%%%%%%%%%%%%%%%%%%%%
 \section{\bf Introduction}
 \setcounter{equation}{0}
 %%%%%%%%%%%%%%%%%%%%%%%%%%%%%%%%%%%%%%%%%%%%%%%%%%%%%%%%%%%%%%%%%%%%%%%%

   In this paper we investigate  composite electric
 $S$-brane solutions (space-like analogues of D-branes) in an
 arbitrary number of dimensions $D$ with a set of scalar fields and
 $(p+2)$-form. The $(p+2)$-form is considered using a composite
 electric ansatz and the metric is taken as diagonal. All ansatz
 functions for the metric, form field and scalar field are taken to
 be dependent upon one distinguished coordinate $u$.
 (For $S$-brane solutions see
  \cite{BF}-\cite{IS} and references therein.)
 In \cite{IS} a  special case of this model containing one scalar field
 with positive-definite kinetic term was considered.

 Here,  as in \cite{IS}, we investigate our system
 similar to the approach used in \cite{IMC}.
 We use the same sigma-model approach with constraints
 and use the result of the previous work \cite{IS},
 where it was  shown that it is possible to satisfy the
 constraints maximally ({\it i.e.} all the branes may carry non-zero
 charge densities) in certain odd dimensions $D = 4m+1 = 5, 9, 13, ...$.

 Here we also examine  cosmological-type solutions in these odd dimensional cases, and discuss
 some of their interesting features such as their Kasner-like
 behavior, existence of non-singular (e.g. bouncing) solutions and
 solutions with acceleration among them.

 In the next two sections we will give the set up for the system
 of $D=n+1$ dimensional gravity with a  set of scalar
 fields and $(p+2)$-form field. For the conditions considered in
 this paper (diagonal metric, composite $Sp$-brane ansatz for the
 antisymmetric $(p+2)$-form field, and all the ansatz functions
 depending only on one coordinate) this  system can be
 reduced to a 1-dimensional $\sigma$-model with contstraints.
 %This $\sigma$-model representation helps in studying the
 %solutions of the system.

 In section   4 we consider the quadratic constraints for the charge
 densities of the branes. Here we briefly remind one of the main
 result of \cite{IS} where it was found
 that these constraints have ``maximal''
 solutions with all non-zero brane charge densities  in particular odd
 dimensions with particular ranks of form fields: $D= 4m +1$ and $p = 2m -1$,
 respectively ($m =1, 2, \dots$).

 In  section 5 we investigate
 cosmological solutions to the field equations for these odd
 dimensions.  We look at the  proper time behavior of the
 simplest of these solutions. We  show that certain solutions
 exhibit Kasner-like behavior at the early times, and some them
 have also  such behavior for big times. Also
 we find an accelerated expansion of $n$-dimensional space
 for certain negative values of dilatonic vector coupling squared,
 i.e. when there at least one ``phantom'' scalar field in the
 model.

 In  section 6  we single out two subclasses of non-singular
 solutions, e.g.   with bouncing behavior of the scale factors.
 Here we also show that solutions with bouncing take place only
 when there is at least one  ``phantom''
 scalar field in the model.

 %%%%%%%%%%%%%%%%%%%%%%%%%%%%%%%%%%%%%%%%%%%%%%%%%%%%%%%%%%%%%%%%%%%%%%%%
  \section{\bf D-dimensional gravity coupled to $q$-form and scalar field}
  \setcounter{equation}{0}
 %%%%%%%%%%%%%%%%%%%%%%%%%%%%%%%%%%%%%%%%%%%%%%%%%%%%%%%%%%%%%%%%%%%%%%%%

 Here  we consider the model governed by the action
   \beq{2.1i}
    S =
       \int_{M} d^{D}z \sqrt{|g|} \left[ {R}[g] -
    h_{ab}  g^{MN} \partial_{M} \varphi^a \partial_{N} \varphi^b
    -  \frac{1}{q!} \exp( 2 \lambda_a \varphi^a ) F^2 \right],
   \eeq
 where $g = g_{MN} dz^{M} \otimes dz^{N}$ is the metric,
 $\vec{\varphi} = (\varphi^a)$   is a set
 (vector) of scalar fields,
 $a = 1, \ldots, l$,  $\vec{\lambda} = (\lambda^a) \in  \R^l$ is a
 constant vector (set) of dilatonic couplings,
 $(h_{ab})$ is a symmetric non-degenerate $l \times l$ matrix
 and
   \beq{2.2i}
   F =  dA =
   \frac{1}{q!} F_{M_1 \ldots M_{q}}
   dz^{M_1} \wedge \ldots \wedge dz^{M_{q}} ,
   \eeq
 is a $q$-form, $q =  p +2 \geq 2$, on a $D$-dimensional manifold
 $M$. Without loss of generality the matrix $(h_{ab})$ may be
 considered as a diagonal one.

 In (\ref{2.1i}) we denote $|g| = |\det (g_{MN})|$ and
   \beq{2.3i}
   F^2 =
         F_{M_1 \ldots M_{q}} F_{N_1 \ldots N_{q}}
         g^{M_1 N_1} \ldots g^{M_{q} N_{q}}.
   \eeq

 The equations of motion corresponding to  (\ref{2.1i}) are
  following
   \bear{2.4i}
   R_{MN} - \frac{1}{2} g_{MN} R  =   T_{MN},
   \\
   \label{2.5i}
   {\btu}[g] \varphi^a -  \frac{\lambda^a}{q!}
    e^{2 \vec{\lambda} \vec{\varphi} } F^2 = 0,
   \\
   \label{2.6i}
   \nabla_{M_1}[g] (e^{2 \vec{\lambda} \vec{\varphi}}
    F^{M_1 \ldots M_{q}})  =  0,
   \ear
 where $\lambda^a = h^{ab} \lambda_b$,
 $(h^{ab})$ is a  matrix inverse to $(h_{ab})$ and
 $\vec{\lambda} \vec{\varphi} = \lambda_a \varphi^a$.

 In (\ref{2.5i}) and (\ref{2.6i}), ${\btu}[g]$ and ${\btd}[g]$ are
 Laplace-Beltrami and covariant derivative operators corresponding
 to  $g$. Equations (\ref{2.4i}), (\ref{2.5i}) and (\ref{2.6i})
 are,  respectively, the multidimensional Einstein-Hilbert
 equations, the "Klein-Fock-Gordon" equation for the scalar field
 and the "Maxwell" equations for the $q$-form.

 The stress-energy tensor in (\ref{2.4i}) can be split up as
   \bear{2.7i}
   T_{MN} =   T_{MN}[\varphi,g]
   + e^{2 \vec{\lambda} \vec{\varphi}} T_{MN}[F,g],
   \ear
   with
   \bear{2.8i}
   T_{MN}[\vec{\varphi},g] =
   h_{ab}
   (\p_{M} \varphi^a \p_{N} \varphi^b -
   \frac{1}{2} g_{MN} \p_{P} \varphi^a \p^{P} \varphi^b),
   \\
   T_{MN}[F,g] = \frac{1}{q!} \left[ - \frac{1}{2} g_{MN} F^2
   + q  F_{M M_2 \ldots M_{q}} F_{N}^{~ M_2 \ldots M_q}\right] ,
   \label{2.9i}
   \ear
 being the stress-energy tensor of scalar fields and $q$-form,
 respectively.

 Let us consider the manifold
   \beq{2.10g}
    M = (u_{-}, u_{+})  \times \R^{n}
   \eeq
 with the diagonal metric
   \beq{2.11g}
       g= w \e^{2{\gamma}(u)} du \otimes du +
    \sum_{i= 1}^{n} \e^{2 \phi^i(u)} \eps_i dy^i \otimes dy^i ,
   \eeq
 where $w=\pm 1$, and $u$ is a distinguished coordinate. The
 functions $\gamma (u), \phi^i (u)$, the scalar fields
 $\varphi^a (u)$ and the $q$-forms are assumed to depend only on
 $u$.

 Here
   \beq{2.14g}
    \eps_i = \pm 1
   \eeq
 are signature parameters,  $i= 1, \ldots, n$.
 When $w = -1$  and all $\eps_i =  1$
 the solutions are cosmological.   The functions
 $\gamma,\phi^i$: $(u_-,u_+) \to \R$ are smooth.

  Here as in \cite{IS}  we define
   \beq{2.15g}
   \Omega_0 = \{ \emptyset, \{ 1 \}, \ldots, \{ n \},
              \{ 1,2 \}, \ldots, \{ 1,  \ldots, n \} \}
   \eeq
 which is the set of all subsets of
   \beq{2.25n}
   I_0 \equiv\{ 1, \ldots, n \}.
   \eeq
 %The set  $\Omega_0$ describes  numbers and ranges of  indices for
 %antisymmetric forms on $M$.

  For any $I = \{ i_1, \ldots, i_k \} \in \Omega_0$ with
  $i_1 < \ldots <  i_k$, we define a form of rank $d(I) \equiv  k$
   \beq{2.17i}
   \tau(I) \equiv dy^{i_1}  \wedge \ldots \wedge dy^{i_k},
   \eeq

 The corresponding brane submanifold is described by coordinates
  $y^{i_1},   \ldots,  y^{i_k}$. We also define the ${\cal E}$-symbol as
    \beq{2.19e}
          {\cal E}(I) \equiv  \eps_{i_1} \ldots \eps_{i_k}.
    \eeq

 Here we adopt the following electric composite $Sp$-brane ansatz for
 the field of the $(p+2)$-form
   \beq{2.27n}
       F = \sum_{I \in \Omega_{e}} d \Phi^{I} \wedge \tau(I)
   \eeq
    where the set
   \beq{2.d1}
    \Omega_{e} \equiv \{I \in \Omega_{0}|  d(I) = q - 1 = p + 1\}
   \eeq
 contains all subsets of $\Omega_0$ of the ``length'' $p+1$, {\it
 i.e.} of the form $\{ i_0, i_1, ..., i_p \}$.

 We assume that the scalar potential and all scalar fields
 depend on the distinguished coordinate
    \beq{2.28nn}
       \Phi^I  = \Phi^I(u) ~, ~~  \varphi^a = \varphi^a(u).
    \eeq

   %%%%%%%%%%%%%%%%%%%%%%%%%%%%%%%%%%%%%%%%%%%%%%%%%%%%%%%%%%%%%%%%%%%%%
    \section{\bf $\sigma$-model representation with constraints}
    \setcounter{equation}{0}
   %%%%%%%%%%%%%%%%%%%%%%%%%%%%%%%%%%%%%%%%%%%%%%%%%%%%%%%%%%%%%%%%%%%%%

    \subsection{\bf $\sigma$-model}

  As it was shown in \cite{IMC} (see Proposition 2 in \cite{IMC})
 the diagonal part of Einstein  equations (\ref{2.4i})
 and the equations of motion (\ref{2.5i})--(\ref{2.6i}), for the
 ansatz given in (\ref{2.11g}), (\ref{2.27n})--(\ref{2.28nn}), are
 equivalent to the equations of motion for a 1-dimensional
 $\sigma$-model with the action (see also \cite{GrIM,IMJ})
   \beq{2.25gn}
    S_{\sigma} =
     \frac{1}{2} \int du {\cal N} \left[ G_{ij} \dot \phi^i \dot \phi^j
     +  h_{ab} \dot \varphi^{a} \dot \varphi^{b}
       + \sum_{I \in \Omega_{e}}
     {\cal E}(I) \exp[-2U^I(\phi, \vec{\varphi})](\dot\Phi^I)^2
   \right],
   \eeq
 the overdots represent differentiation with respect to $u$,
 {\it i.e.} $\frac{d}{du}$.

 The factor ${\cal N}$ is the lapse function given by
   \beq{2.24gn1}
     {\cal N}= \exp(\gamma_0-\gamma)>0
   \eeq
 where
     \beq{2.24gn}
     \gamma_0(\phi)   \equiv \sum_{i=1}^n \phi^i,
    \eeq
 The factor in the exponent is given by
     \beq{2.u}
       U^I = U^I(\phi,\vec{\varphi})= - \vec{\lambda} \vec{\varphi} +
                               \sum_{i \in I} \phi^i.
     \eeq
 Here and it what follows
   \beq{2.c}
    G_{ij}= \delta_{ij}- 1
   \eeq
 are components of the ``pure cosmological'' minisupermetric matrix,
 $i,j= 1, \dots, n$, see  \cite{IM2,IMZ}.

 The generalized ``Maxwell equations'' of (\ref{2.6i})
   \beq{5.29n}
    \frac d{du}\left(\exp(-2U^I) \dot \Phi ^I \right)=0
   \eeq
 can be readily  integrated to give
   \beq{5.29na}
    \dot \Phi ^I= Q(I) \exp(2U^I),
   \eeq
 where $Q(I)$ are constant charge densities,  $I \in \Omega_{e}$.

 We will fix the time gauge to be harmonic one
   \beq{4.1n}
        \gamma= \gamma_0,  \quad  {\cal N} = 1.
   \eeq

 We also introduce collective variables $x=(x^A)=(\phi^i,\varphi^a)$ and a
 ``truncated'' target space metric
    \bear{2.35n}
     \bar G=\bar G_{AB}dx^A\otimes dx^B=
     G_{ij}d\phi^i\otimes d\phi^j+
    h_{ab} d\varphi^a \otimes d\varphi^b, \\ \label{2.36n}
     (\bar G_{AB})=\left(\begin{array}{cc}
     G_{ij}&0\\
     0& h_{ab}
    \end{array}\right).
    \ear
 The $U^I$-vectors  defined in (\ref{2.u}) can be written as $U^I(x)=U_A^I
 x^A$ with components
    \beq{2.38n}
      (U_A^I)=(\delta_{iI},- \lambda_{a})~,
   \eeq
 where
    \beq{2.39n}
    \delta_{iI}\equiv \sum_{j\in I} \delta_{ij}=
    \begin{array}{ll}
                     1, &i \in I; \\
                     0, &i \notin I;
    \end{array}
    \eeq
 is an indicator of $i$ belonging to $I$. For fixed charge
 densities $Q(I)$, $I \in \Omega_{e}$, the equations of motion for
 the $\sigma$-model  in (\ref{2.25gn}) are  equivalent to the
 Lagrange equations corresponding to the Lagrangian
   \beq{5.31n}
     L_Q=\frac12 \bar G_{AB} \dot x^A\dot x^B-V_Q,
   \eeq
 with the zero-energy constraint
   \beq{5.33n}
     E_Q=\frac12 \bar G_{AB} \dot x^A \dot x^B + V_Q = 0
   \eeq
imposed.
 Here
   \beq{5.32n}
     V_Q= \frac12\sum_{I \in \Omega_{e}}{\cal E}(I) Q^2(I) \exp[2U^I(x)].
   \eeq

 In section 5 we will find explicit solutions to the field
 equations  resulting from (\ref{5.31n}) - (\ref{5.32n})
 for certain dimensions $D = 4m +1$ and positive $\eps_i$.

   %%%%%%%%%%%%%%%%%%%%%%%%%%%%%%%%%%%%%%%%%%%%%%%%%%%%%%%%%%%%
    \subsection{Constraints}
   %%%%%%%%%%%%%%%%%%%%%%%%%%%%%%%%%%%%%%%%%%%%%%%%%%%%%%%%%%%%

 Due to diagonality of the Ricci-tensor for the metric
 (\ref{2.11g}) the non-diagonal part of the Einstein equations
 (\ref{2.4i}) reads as follows
   \beq{2.4ij}
     T_{i j} = 0, \qquad i \neq j.
   \eeq

 This leads to constraints on the charge densities $Q(I)$.
 The non-diagonal components of stress-energy tensor are
 proportional to
    \beq{2.4e}
     e^{ 2 \vec{\lambda} \vec{\varphi}} F_{i M_2 \dots M_{q}} F_{j}^{~ M_2 \dots M_q},
    \eeq
 with $i \neq j$. From (\ref{2.27n}) (\ref{2.u}) and (\ref{5.29na})
 we obtain for the $(p+2)$-form
    \beq{2.27nn}
    F=   \frac{1}{(p+1)!} Q_{i_0 \dots i_p}
       \exp(2 \phi^{i_0} + \dots + 2 \phi^{i_p} -
       2 \vec{\lambda} \vec{\varphi})
       du \wedge d y^{i_0} \wedge \dots \wedge d y^{i_p}
   \eeq
 Inserting this in (\ref{2.4e}) we are led to the following
 constraint equations on charge densities \cite{IMC}
    \beq{2.4c}
     C_{ij} \equiv  \sum_{i_1, \dots, i_p =1}^{n}
     Q_{i i_1 \dots i_p} Q_{j i_1 \dots i_p} \eps_{i_1} \e^{2 \phi^{i_1}}
     \dots   \eps_{i_p} \e^{2 \phi^{i_p}} = 0,
    \eeq
 where $i \neq j$; $i, j =1, \dots,n$. ($T_{i j}$ is proportional to
 $\exp(- 2 \vec{\lambda} \vec{\varphi} - 2 \gamma + 2 \phi^i + 2 \phi^j)
  C_{i j}$ for $i \neq j$.)

 Here $p = q - 2$  and $Q_{i_0 i_1 .. i_p}$ are components of the
 antisymmetric form of rank $p+1 = q-1$ and
    \beq{2.4d}
     Q_{i_0 i_1 \dots i_p} =  Q(\{ i_0, i_1, \dots , i_p \})
    \eeq
 for $i_0 < i_1 < \dots < i_p$ and $\{ i_0, i_1, \dots , i_p \} \in
 \Omega_e$.

 The number of constraints in (\ref{2.4c}) is $n(n -
 1)/2$. It was shown in \cite{IS} that these constraints
 can  be satisfied when the dimension of space-time takes certain
 odd values (see next section).

   %%%%%%%%%%%%%%%%%%%%%%%%%%%%%%%%%%%%%%%%%%%%%%%%%%%%%%%%%%%%%%%%%%%%%%%%
   \section{Solution to constraints for  D= 4m+1}
   %%%%%%%%%%%%%%%%%%%%%%%%%%%%%%%%%%%%%%%%%%%%%%%%%%%%%%%%%%%%%%%%%%%%%%%%

  As in \cite{IS} we rewrite the  constraints (\ref{2.4c}) as
    \beq{3.c}
   \bar{C}_i^j =  \sum_{i_1, \dots, i_p =1}^{n}
     \bar{Q}_{i i_1 \dots i_p} \bar{Q}^{j i_1 \dots i_p} = 0,
    \eeq
 $i \neq j$; $i, j =1, \dots , n$, where we introduce ``running'' charge densities
     \beq{3.4d}
     \bar{Q}_{i_0 i_1 \dots  i_p} =  Q_{i_0 i_1 \dots i_p}
     \prod_{k = 0}^{p} \exp(\phi^{i_k}).
    \eeq
 Here the indices are lifted by the flat metric
    \beq{3.e}
    \eta = \eps _1 dy^1 \otimes dy^1 + \dots + \eps_n dy^n \otimes
    dy^n = \eta _{ab} dy^a \otimes dy^b,
    \eeq
 where $(\eta _{ab}) = (\eta ^{ab}) =
  {\rm diag}(\eps _1 , \dots , \eps _n)$,
  {\it i.e.} $\bar{Q}^{i_0  \dots i_p} = \eta^{i_0 i_0 '} \dots
  \eta^{i_p i_p '} \bar{Q}_{i_0' \dots i_p'}$.
   Here we denote $\bar{C}_i^j =
   C_i^j \exp(\phi^{i} + \phi^{j})$ with $C_i^j = C_{ik} \eta^{kj}$.

 It was shown in \cite{IS} that when the total space-time dimension is
 the following one
      \beq{3.h}
         D= n+1 = 2(p+1) + 1 = 4m + 1= 5, 9, 13, \dots
      \eeq
   with
     \beq{3.ed}
       p = 2m -1 = 1, 3, 5, \dots
     \eeq
  and
     \beq{3.e1}
      \eps _1 \dots \eps_n =1
     \eeq
 then any  self-dual or anti-self-dual
 ``running'' charge density form
    \beq{3.eab}
    \bar{Q}_{i_0  \dots i_p} =
     \pm \frac{1}{(p+1)!} \eps _{i_0  \dots i_p j_0 \dots j_p}
     {\bar Q}^{j_0 \dots j_p} = \pm (* \bar{Q})_{i_0  \dots i_p}.
     \eeq
 satisfies the constraint equations (\ref{3.c}) (or, equivalently, eqs. (\ref{2.4c})).
 Here the symbol $*=*[\eta ]$ is the Hodge operator with respect to $\eta$.
 The relation (\ref{3.e1}) means that the number of time-like coordinates is even.

 %The dimension of the space of solutions is
 %$\frac{1}{2}C_{2(p+1)}^{p +1}$. The factor of $\frac{1}{2}$ comes from
 %(anti-)self-duality condition.

 {\bf Remark.} It  was shown in \cite{IS} that in the special case:
 $D=5$, $p = 1$ and $\eps _1 = \dots = \eps _4 =1$
 the solutions presented above are the only solutions to
 constraints (\ref{3.c}) when all $\bar{Q}_{i j}$ ($i < j$)
 are imposed to be non-zero ones.

 %%%%%%%%%%%%%%%%%%%%%%%%%%%%%%%%%%%%%%%%%%%%%%%%%%%%%%%%%%%%%%%%%%%%%%%%
  \section{Cosmological-type solutions for  D= 4m+1}
 %%%%%%%%%%%%%%%%%%%%%%%%%%%%%%%%%%%%%%%%%%%%%%%%%%%%%%%%%%%%%%%%%%%%%%%%

  Here we consider cosmological  type solutions   for the dimensions
  from (\ref{3.h}) when all $y^i$-coordinates are space-like,
  {\it i.e.}
    \beq{4.eps}
          \eps _1 = \dots = \eps _n =1.
    \eeq

  It was shown in \cite{IS} that the definition of running constants
    (\ref{3.4d}) and the (anti-) self-duality of the charge
    density form (\ref{3.eab}) imply that all scale factors are the same
      up to constants:
      \beq{4.ba}
       \phi^i(u) = \phi(u) + c^i.
      \eeq

     Without loss of generality
     we put  $c^i = 0$ which may always be done via a proper
     rescaling of $y$-coordinates. This also implies that non-running
     charge density form  $Q_{i_0  \dots i_p}$
     is self-dual or anti-self-dual in a flat
     Euclidean space $\R^n$, i.e.
     \beq{3.eaa}
      Q_{i_0  \dots i_p} =
     \pm \frac{1}{(p+1)!} \eps _{i_0  \dots i_p j_0 \dots j_p}
      Q^{j_0 \dots j_p} = \pm (* Q)_{i_0  \dots i_p}.
     \eeq

     The Lagrangian and total energy constraint are given by
    \bear {4.d}
     L_Q &=& \frac{1}{2}  G_{ij} \dot \phi^i\dot \phi^j-V_Q + \frac{1}{2}
      h_{ab}  \dot \varphi^a \dot \varphi^b , \\
     E_Q &=& \frac{1}{2}  G_{ij} \dot \phi^i \dot \phi^j + V_Q +
     \frac{1}{2}  h_{ab}  \dot \varphi^a \dot \varphi^b  = 0,
     \ear
     with the potential being
     \beq{4.e}
       V_Q = \frac{1}{2} \sum _I Q^2 (I) \exp \left(2 \sum _{k \in I} \phi^k
         - 2 \vec{\lambda} \vec{\varphi} \right),
     \eeq
   see (\ref{5.31n})-(\ref{5.32n}).

  The field equations for $\phi$ and $\varphi^a$ from $L_Q$ are
  the following ones
     \bear{4.f}
     \sum _{j= 1} ^n G_{ij} \ddot \phi ^j +
     \sum _I Q^2 (I) \delta ^i _I \exp
     \left( 2 \sum _{k \in I} \phi ^k -2 \vec{\lambda} \vec{\varphi} \right)
     =  0, \\
     \label{4.fa}
     \ddot \varphi^a + \sum _I Q^2 (I) (-\lambda^a) \exp
     \left( 2 \sum _{k \in I} \phi ^k -2 \vec{\lambda} \vec{\varphi} \right)
     =     0,
     \ear
    $a = 1, \ldots, l$.

   Since all $\phi ^i = \phi$
   we get $\sum _{j=1} ^n G_{ij} \ddot \phi ^j= \sum _{j=1} ^n
   (\delta _{ij} -1) \ddot \phi ^j = (1-n) \ddot \phi$.

   As in \cite{IS}, defining
    \beq{4.q1}
        Q^2  \equiv \sum _I Q^2 (I)  \neq 0
    \eeq
   and noting that due to (\ref{3.eaa})
     \beq{4.q2}
     \sum _I Q^2 (I) \delta _I^i =   \frac{1}{2} Q^2
     \eeq
    for any $i = 1, ... ,n$,  we could write the field equations
    (\ref{4.f}) and (\ref{4.fa}) in the  following form
     \bear{4.g}
     \ddot \phi &=& \frac{1}{2(n-1)} Q^2
           \exp (n \phi - 2 \vec{\lambda} \vec{\varphi}), \\
     \label{4.ga}
     \ddot \varphi^a &=& \lambda^a Q^2 \exp (n \phi - 2 \vec{\lambda}
     \vec{\varphi}),
     \ear
     $a = 1, \ldots, l$.

     Equations (\ref{4.g}) and (\ref{4.ga}) imply
     \beq{4.h}
     \ddot f = -2 A e^{2 f},
     \eeq
     where the definitions
     \bear{4.ha}
     f &\equiv& \frac{n}{2} \phi - \vec{\lambda} \vec{\varphi}, \\
     A &\equiv& \frac{Q^2}{2} K, \qquad
           K \equiv  \vec{\lambda}^2 - \frac{n}{4(n-1)} ,
     \label{4.hb}
      \ear
     are assumed. Here and it what follows
     $\vec{\lambda}^2 = h_{ab}\lambda^a \lambda^b$.

     The first integral of (\ref{4.h}) is
     \beq{4.hc}
       \frac{1}{2} \dot f^2 + A e^{2 f} = \frac{1}{2} C,
     \eeq
     where $C$ is an integration constant.

     Let $K \neq 0$, or
     \beq{lambda}
     \vec{\lambda}^2 \neq  \frac{n}{4(n-1)} \equiv \lambda^2_0.
     \eeq

     Equation (\ref{4.h}) has  several solutions \cite{IS}:
     \beq{4.i}
        f = - \ln \left[ z |2 A|^{1/2} \right]
     \eeq
     with
     \bear{4.ja}
     z &=& \frac{1}{\sqrt{C}} \sinh \left[ (u-u_0) \sqrt{C} \right],
           \qquad A<0 , ~C>0; \\
                          \label{4.jb}
     &=& \frac{1}{\sqrt{-C}} \sin \left[ (u-u_0) \sqrt{-C} \right],
           \qquad A<0 , ~C<0; \\
                           \label{4.jc}
     &=& u-u_0, \qquad \qquad \qquad \qquad \qquad A<0, ~C=0; \\
                            \label{4.jd}
      &=& \frac{1}{\sqrt{C}} \cosh \left[ (u-u_0) \sqrt{C} \right],
     \qquad A>0 , ~C>0.
     \ear

     Using (\ref{4.g}) and (\ref{4.ga} we get the
     following relationship
     \beq{4.l}
     \ddot \varphi^a = 2 (n-1) \lambda^a \ddot \phi,
     \eeq
     which has the solution
     \beq{4.m}
      \varphi^a = 2 (n- 1) \lambda^a \phi + C_2^a u + C_1^a,
     \eeq
     where $C_2^a , C_1^a$ are integration constants,
     $a = 1, \ldots, l$.
     Combining
     (\ref{4.m}) with (\ref{4.ha}) gives
     \beq{4.n}
     \phi = \frac{1}{2 (1- n) K} \left[f(u) +
     \vec{\lambda}(\vec{C}_2 u + \vec{C}_1) \right],
     \eeq
     and
      \beq{4.ns}
      \varphi^a = - \frac{\lambda^a}{K}
      \left[ f(u) + \vec{\lambda}(\vec{C}_2 u + \vec{C}_1) \right]
       +  C_2^a u + C_1^a,
      \eeq
      $a = 1, \ldots, l$.

     Substituting this to the zero-energy constraint
     \beq{4.o}
     E_Q = \frac{1}{2} n (1-n) \dot \phi ^2 +
     \frac{1}{2} h_{ab} \dot \varphi^a \dot \varphi^b +
     \frac{1}{2} Q^2 e ^{2 f(u)} = 0
     \eeq
     and using  (\ref{4.hc}) we get
     \beq{4.o1}
      2E_Q = \frac{C}{K} + (\vec{C}_2)^2
           - \frac{ (\vec{\lambda} \vec{C}_2)^2}{K} =  0,
     \eeq
     or, equivalently,
     \beq{4.o2}
     C = (\vec{\lambda} \vec{C}_2)^2 - K (\vec{C}_2)^2 =
      (\vec{\lambda} \vec{C}_2)^2 - \vec{\lambda}^2
      (\vec{C}_2)^2 + \frac{n}{4(n-1)}(\vec{C}_2)^2.
     \eeq
     where $(\vec{C}_2)^2 = h_{ab} C_2^a C_2^b$ and
      $\vec{\lambda} \vec{C}_2 = \lambda_a C_2^a$.

     Thus, the solutions for the
     metric, scalar fields and $(p + 2)$-form read:
     \bear{4.pa}
     ds^2 &=& we^{2n \phi(u)} du^2 + e^{2 \phi(u)}
     \sum _{i=1}^n (dy^i)^2 \\
     \label{4.pb}
     \varphi^a &=& - \frac{\lambda^a}{K}
      \left[ f(u) + \vec{\lambda}(\vec{C}_2 u + \vec{C}_1) \right]
       +  C_2^a u + C_1^a, \\
     \label{4.pc}
     F &=& e^{2 f(u)} du \wedge Q,
     \qquad
     Q =  \frac{1}{(p+1)!} Q_{i_0  \dots i_p}
                           dy^{i_0}  \wedge \dots  \wedge dy^{i_p},
     \ear
     $a = 1, \ldots, l$,
     with  $\phi (u)$ given by (\ref{4.n}) and the function
     $f(u)$ given by (\ref{4.i}) and (\ref{4.ja}),
     (\ref{4.jb}), (\ref{4.jc}), (\ref{4.jd}).
     Here the charge density form  $Q$
     of rank $n/2 = 2m$  is self-dual or anti-self-dual in a flat
     Euclidean space $\R^n$:  $Q = \pm * Q$,
     the parameters $C_2^a, C$ obey (\ref{4.o2}) and the vector of
     dilatonic coupling constants $\vec{\lambda}$ is non-exceptional
     one,  see (\ref{lambda}).

    \subsection{Special attractor solution for $C_2^a = 0$ }

 Here we consider the simplest
 cosmological type solution  with $C_2^a = 0$
 and, hence, $C = 0$, i.e.  the solution
 corresponding to (\ref{4.jc}).  For this
 solution $A < 0$ and hence
      \beq{5.l}
        \vec{\lambda}^2 < \lambda^2_0.
      \eeq
 Without loss of generality we put $u_0 =  0$.

 Let us introduce a ``proper time'' coordinate:

     \beq{5.t}
     d\tau = -  e^{n \phi (u)} du,
     \eeq
     where
     \beq{5.ph}
     \phi = \frac{1}{2 (n - 1) K} \ln (u |2 {\bar A}|^{1/2}),
     \qquad
     {\bar A} = A e^{- 2 \lambda_a C_1^a}.
     \eeq
 Integrating  (\ref{5.t}) and taking a suitable choice of reference
 point we get for $\vec{\lambda}^2 \neq - \lambda^2_0$
     \beq{5.tau}
     |\alpha | |2 {\bar A}|^{1/2} \tau =
     (u |2 {\bar A}|^{1/2})^{\alpha},
     \eeq
     where $u >0$ and
     \beq{5.al}
     \alpha = \frac{\vec{\lambda}^2 +
     \lambda^2_0}{\vec{\lambda}^2 - \lambda^2_0}.
     \eeq

 We note that for $\vec{\lambda}^2 > - \lambda^2_0$ we get
 $\alpha <0$, and hence $\tau = \tau (u)$ is monotonically decreasing
 from infinity (when $u=0$) to zero (when $u=\infty$).

 The metric  (\ref{4.pa}) now reads
     \beq{5.pa}
       ds^2 = w d \tau^2 + B \tau^{2 \nu} \sum _{i=1}^n (dy^i)^2,
     \eeq
 where $\tau > 0$ and
     \beq{5.1a}
       \nu = \frac{2}{n + 4 \vec{\lambda}^2 (n-1)}, \qquad
        B = (|\alpha| |2 {\bar A}|^{1/2})^{2 \nu}.
     \eeq

 For scalar field we get the following relations
    \beq{5.pb}
      \varphi^a = 2 \lambda^a (n-1) \nu \ln \tau  +  \varphi^a_0,
    \eeq
      $a = 1, \ldots, l$; where $B = B_0 \exp(- 4 \lambda_a \varphi^a_0)$
      and $B_0 = (|\alpha| |2 A|^{1/2})^{2 \nu}$.

 For $(p+2)$-form we get
      \beq{5.pc}
      F =  \frac{(d\tau \wedge Q)}{2 \alpha |A| \tau}
            \frac{|2A|^{(\alpha -1)/2 \alpha}}{(|\alpha|
            \tau)^{1/\alpha}} \exp(- 2 \lambda_a \varphi^a_0).
       \eeq

 By putting $\vec{\lambda} =0$ and $w = -1$ in the above
 solution  we get a cosmological power-law expansion with a power
 $\nu = 2/n$ that is the same as in the case of $D=1+n$ dust matter with a
 zero pressure (see, for example \cite{Lor1,BO,IM2}). This coincidence was
 explained in \cite{IS}, i.e. it was shown that  the collection of
 branes with charge densities obeying (anti)-self-duality
 condition (\ref{3.eaa}) behaves as a dust matter.

 The solution given by relations
 (\ref{5.pa}), (\ref{5.pa}),(\ref{5.pc}) is an attractor
 solution in the limit  $\tau \to + \infty$,
 or $u \to + 0$,  for the  solutions
 with $A < 0$  given by (\ref{4.ja}). This follows
 just from the relation $\sinh u  \sim u$ for small $u$.

  {\bf Accelerated expansion.} When
   \beq{5.ac}
       \nu > 1
   \eeq
 we get an accelerated expansion of $n$-dimensional space. This
 relation is equivalent to the following one $(n > 2)$
   \beq{5.acc}
        - \frac{n}{4(n-1)} < \vec{\lambda}^2 < - \frac{(n -
        2)}{4(n-1)}.
   \eeq
 This means that $\vec{\lambda}^2 < 0$ and, hence, the matrix
 $(h_{ab})$ is not a  positive-definite one. So, there should be at
 least one scalar field  with negative kinetic term (i.e. so-called
 phantom field).

    \subsection{Kasner-like behavior for $\tau \to + 0$ }

  Now, we consider $u \to + \infty$ asymptotical behavior of solutions
 with i) $\sinh$- and ii) $\cosh$- functions corresponding to
 (\ref{4.ja}) and (\ref{4.jd}), respectively.
 In both cases we find
 that the asymptotic behavior is Kasner like.
 We remind that in general case
 such behaviour is described by the following
 asymptotical relations for the metric
 and scalar fields \cite{IMb1,DIMb}
     \begin{equation}
         ds^2_{as} = w d\tau^2 +
        \sum_{i=1}^{n}\tau^{2 \alpha ^i} A_i (dy^i)^2,
       \qquad
       \varphi^a_{as} =  \alpha_{\varphi}^a \ln \tau + \varphi_0^a,
 \end{equation}
 where $A_i > 0$,  $\varphi_0^a$ are constants. The Kasner
 parameters obey

  \begin{equation} \label{5.4}
   \sum _{i=1}^n \alpha^i = \sum _{i=1}^n
   (\alpha ^i) ^2 + h_{ab}\alpha _{\varphi}^a \alpha _{\varphi}^b = 1.
  \end{equation}

 In our case all $\alpha^i$ are coinciding since we have an isotropic
 expansion described by the metric.

 Here we consider
 for simplicity the case when the matrix $(h_{ab})$ is positive
 definite one. In what follows we consider solutions with $C > 0$.
 We also put $\vec{C}_2 \neq 0$ that imply
  $(\vec{C}_2)^2 > 0$.

 In the first case i) with $\sinh$-dependence and $K < 0$
 (or, equivalently, $\vec{\lambda}^2 < \lambda^2_0$)
 the proper time $\tau$ is decreasing when $u \to + \infty$. From eqs.
 (\ref{4.i}), (\ref{4.ja}),  (\ref{4.jd}), (\ref{4.n}) and  (\ref{5.t})
 we find the  following asymptotic behavior as $u \to + \infty$
     \bear{5.3}
       &&\phi \sim  - b u +  {\rm const}, \qquad
         b =  \frac{\vec{\lambda} \vec{C}_2 - \sqrt{C}}{2(n -1) K} > 0,
                     \\   \label{5.3a}
       &&\varphi^a \sim [ C_2^a - 2 (n- 1) b \lambda^a ]u + {\rm const},
                     \\   \label{5.3b}
       && \tau \sim {\rm const} \exp( - nb u).
    \ear

The inequality (\ref{5.3}) follows just from (\ref{4.o2}).

 Using these asymptotic relations and writing everything in terms of proper
 time one find that the metric and scalar field take the following
 asymptotic forms
     \bear{5.2}
       &&ds^2_{as} =
       w d\tau^2 + \tau^{2/n} A_0 \sum_{i=1}^{n}(dy^i)^2,
                    \\ \label{5.2a}
       &&\varphi_{as}^a =  \alpha_{\varphi}^a \ln \tau + \varphi_0^a,
          \ear
 as $\tau \to + 0$. Here  $A_0 > 0$,  $\varphi_0^a$ are constants
 and
  \beq{5.2b}
     \alpha_{\varphi}^a  = \frac{2 (n-1)\lambda^a}{n} -
     \frac{C_2^a}{nb}.
  \eeq

 It may be verified that the Kasner relation
   \beq{5.k}
    h_{ab} \alpha_{\varphi}^a \alpha_{\varphi}^b = 1 - \frac{1}{n}
    \eeq
 is satisfied identically (in our case  all  $\alpha ^i =1/n$).

 %Using (\ref{5.3})-(\ref{5.3b}) one can obtain
 %the following relationship:
 %$h_{ab} \alpha_{\varphi}^a C_{2}^b < 0$.

 In the second case ii) with $\cosh$-dependence and $K > 0$
(or, equivalently, $\vec{\lambda}^2 > \lambda^2_0$)
 the
 proper time $\tau$ decreases as $u \to +\infty$ for $\lambda_a C_2^a >
  0$  and increases for $\lambda_a C_2^a < 0$. In this case we also get
 an asymptotical Kasner type relations (\ref{5.2})-(\ref{5.2a}) boths in
 the limits  $\tau \to + 0$ and $\tau \to + \infty$.

 The Kasner parameters for scalar fields have the following form
  \beq{5.alb}
  \alpha_{\varphi, \mp}^a  = \frac{2 (n-1)\lambda^a}{n} -
           \frac{C_2^a}{n b_{\mp}}
   \eeq
   where
 \beq{5.ab}
  b_{\mp} = \frac{\vec{\lambda} \vec{C}_2 \mp
  {\rm sign} (\vec{\lambda} \vec{C}_2) \sqrt{C}}{2(n -1) K}.
   \eeq
 The signs "-" and "+" refer to the asymptotical behaviors  for
 $\tau \to + 0$ and  $\tau \to + \infty$, respectively.

 In both $\sinh-$ and $\cosh-$ cases the  Kasner set
 $\alpha = (\alpha^i = 1/n, \alpha_{\varphi}^a)$ corresponding to
 the limit $\tau \to + 0$  obeys the  inequalities
     \beq{5.u}
        U^I(\alpha)= - \lambda_a \alpha_{\varphi}^a +
                               \sum_{i \in I} \alpha^i
        = - \lambda_a \alpha_{\varphi}^a +  \frac{1}{2} > 0
     \eeq
 for all brane sets $I$. This is in agreement with a general
 prescription of the billiard representation  from \cite{IMb1}.

 In the case ii) we obtain that the Kasner set
 corresponding to  the limit $\tau \to + \infty$
 obeys the  inequalities
     \beq{5.v}
        U^I(\alpha) <   0
     \eeq
  for all $I$.

  Relations (\ref{5.u}) and (\ref{5.v}) are nothing more then
  the  conditions for vanishing of the $q$-form term
  $\exp( 2 \lambda_a \varphi^a ) F^2$  in the Lagrangian
  (\ref{2.1i}) in Kasner regimes with
  $\tau \to + 0$ and  $\tau \to + \infty$, respectively
  (see \cite{IMb1}).

  {\bf Scattering law for Kasner parameters.}
  The Kasner  parameters for "in" ($\tau \to + 0$)
  and "out" ($\tau \to + \infty$) asymptotics
  are related as follows
 \beq{5.sl}
   \alpha_{\varphi, +}^a  = \frac{2 (n-1)\lambda^a}{n}
    - \frac{\alpha_{\varphi, -}^a  - \frac{2 (n-1)\lambda^a}{n}}
     {1+ \frac{n}{(n-1)K}[\lambda_a \alpha_{\varphi, -}^a  - \frac{2
        (n-1)\vec{\lambda}^2}{n}]}.
  \eeq

  This relation could be readily obtained from formulas
   (\ref{5.alb}) and  (\ref{5.ab}).
   Let us explain the geometrical sense of (\ref{5.sl}) putting
   for simplicity $h_{ab} = \delta_{ab}$. Vectors $\vec{\alpha}_{\varphi, -}$
    and $\vec{\alpha}_{\varphi, +}$ belong to the $(l-1)$-dimensional
    sphere of radius $R = \sqrt{1-
    \frac{1}{n}}$.   Let us introduce a vector
    $\vec{\Lambda} = \frac{2 \vec{\lambda} (n-1)}{n}$.
    The endpoint of this  vector is out of the sphere (of radius $R$) due to $K >
    0$. If we consider a line connecting the endpoints of
    the vectors  $\vec{\Lambda}$ and  $\vec{\alpha}_{\varphi, -}$,
    then we get another point of intersection of this line with
    the sphere. This another point corresponds to the vector
    $\vec{\alpha}_{\varphi, +}$.

    {\bf Remark.} In \cite{DIMb} we have suggested an example of
    never-ending oscillating  behavior near the singularity
    supported by scalar fields with multiple exponential
    potential. In \cite{DIMb} we have found a ``reflection law'' for scalar
    Kasner parameters using the general relation from
    \cite{Ierice}. Here we suggest an example of ``Kasner scattering'', when Kasner
    behaviour takes place both   ``in'' ($\tau \to + 0$) and ``out'' ($\tau \to + \infty$)
    asymptotics. In this case the ``scattering'' also takes place
    only for scalar part of Kasner parameters, while Kasner
    parameters for the metric are coinciding in both asymptotics.

    {\bf Remark.} Here as in \cite{IMbpf} we  also deal with the
    illumination of a sphere by a point-like sources.
    Indeed, if we put a source of light into the endpoint of the
    vector $\vec{\Lambda}$ then we get two domains on the sphere:
    i) the domain illuminated (strongly) by the source of light;
    and ii) the  shadow domain. Due to inequalities (\ref{5.u})
    and (\ref{5.v}) the vector $\vec{\alpha}_{\varphi, -}$
    belongs to the shadow domain and the vector $\vec{\alpha}_{\varphi, +}$
    belongs to the illuminated one.     The map (\ref{5.sl}) transforms
    from the shadow domain into another one.
    %The fixed points of
    %the map (\ref{5.sl}), i.e. those obeying
    %$\alpha_{\varphi, +}^a = \alpha_{\varphi, -}^a$,
    %belong to the border between two domains. This border
    %submanifold is a $(l-2)$-dimensional sphere.

    \section{Non-singular solutions }

  Here we single out   subclasses of non-singular solutions, e.g.
  with bouncing behavior of scale factor
        \beq{6.1}
         a = e^{\phi (u)} = \exp \left[\frac{1}{2(1-n)K} \bar{f} (u)  \right],
        \eeq
  where
      \beq{6.2}
          \bar{f} = f + \lambda_a (C_2^a u + C_1^a).
        \eeq

  In this section we put
       \beq{6.2a}
         \vec{\lambda}^2 > - \lambda^2_0.
        \eeq

  Let us  consider the case $K < 0$, or equivalently,
  $\vec{\lambda}^2 < \lambda^2_0$. In what follows we use
  synchronous or proper time variable from (\ref{5.t}), or, equivalently,
       \beq{6.3}
          \tau = \int_{u}^{u_{*}} du'  e^{n \phi (u')}.
        \eeq

   For $C < 0$ we have a  bouncing behavior of scale factor

   It follows from (\ref{6.1}), (\ref{6.2}) and (\ref{6.3}) that the solution
   is a non-singular one for $\tau \in ( -  \infty, + \infty )$
   in such three cases
   \bear{6.4}
      (A_{+}) \quad
               \lambda_a C_2^a  >  \sqrt{C}, \quad  \  C  \geq 0;
              \\ \label{6.5}
      (A_{0}) \quad \quad
              \lambda_a C_2^a  = \sqrt{C}, \quad  \  C  > 0;
              \\ \label{6.6}
      (A_{-}) \qquad \qquad \qquad \qquad   \ C < 0.
    \ear

    Indeed, for $u = u_0$, the integral in (\ref{6.3})
    is   divergent in all three cases due to
      \beq{6.4a}
           e^{n \phi (u)} \sim (u - u_0)^{\alpha - 1},
       \eeq
  as $u \to u_0 + 0$, since $\alpha < 0$ (see, (\ref{5.al}))
  according to (\ref{6.2a}). This implies $\tau(u) \to + \infty$
  as $u \to u_0 + 0$. The relation (\ref{6.4a}) just follows from
  relations $\sinh(x) \sim \sin(x) \sim x $ for small $x$.

 In the case $(A_{-})$ the function $\phi (u)$ is defined on the
 interval $(u_0, u_{+})$ and
   \beq{6.4b}
           e^{n \phi (u)} \sim (u_{+} - u)^{\alpha - 1},
       \eeq
  as $u \to u_{+} -0$. This follows just from $\sin(\pi - x)=
 \sin(x) \sim x$ for small $x$. Relation (\ref{6.4b})
 implies $\tau(u) \to - \infty$   as $u \to u_{+} - 0$.
 Thus, in the case $(A_{-})$ the solution is defined
 for $\tau \in ( -  \infty, + \infty )$.

    For $(A_{+})-$ and  $(A_{0})-$ cases the solution is defined for
    $u \in  (u_0 , + \infty )$.
    When $C > 0$ the function (\ref{6.2}) has the following
    asymptotical behavior
    \beq{6.10a}
    \bar{f} \sim - (u -u_0)\sqrt{C} + \lambda_a C_2^a u + {\rm const}
    \eeq
    for $u \to + \infty$, that implies the divergency
    of the integral in (\ref{6.3}) for $u = + \infty $.
    An analogous consideration could be done for $C =0$
    (with logarithmic term appearing instead of the first term
    in r.h.s. of (\ref{6.10a}) ). Thus, in both cases
    $(A_{+})$ and  $(A_{0})$ the solution
    is defined  for $\tau \in ( -  \infty, + \infty )$.

    In the cases $(A_{+})$ and  $(A_{-})$
    we also get a bouncing behaviour of the scale factor $a(\tau)$
    at some point $\tau_{b}$, i.e.
    the function $a(\tau)$ is monotonically
    decreasing in the interval
    $( -  \infty,  \tau_b )$ and monotonically
    increasing in the interval $( \tau_b, + \infty )$.

    In the case $(A_{0})$  the solution is
    non-singular, the function  $a(\tau)$ is monotonically
    increasing to infinity from a non-zero  value.

    %For other values of parameters the solution is
    %singular in general position (excepting certain
    %special cases),  the function $a(\tau)$
    %is monotonically increasing to infinity
    %from  zero value. This   takes place when either
    %      (i) $\lambda_a C_2^a  <  \sqrt{C}$, $C  > 0$
    %   or
    %      (ii) $C = 0$,  $ \lambda_a C_2^a \leq 0$.
    %The solution for
    %these values of parameters is defined in the
    %semi-infinite interval $( 0, + \infty )$.

    {\bf Remark.} A similar behavior takes place in $D$-dimensional model with  two
    sets of scalar  fields: ``usual'' $\vec{\varphi}$ and phantom $\vec{\psi}$
    ones   and exponential potential depending  upon
    $\vec{\varphi}$, see  \cite{AIM}.

    Let us show that the bouncing behavior for $K <0$ takes place
    only when the matrix $(h_{ab})$ is not positive definite, i.e.
    when at least one scalar field is a "phantom" one.
    Indeed, in the case $(A_{+})$  we get squaring (\ref{6.4})
     \beq{6.7}
           (\lambda_a C_2^a)^2  >  C =(\vec{\lambda} \vec{C}_2)^2 - K
           (\vec{C}_2)^2,
     \eeq
  and, consequently, $(\vec{C}_2)^2 = h_{ab} C_2^a C_2^b < 0$.
  The same inequality we get in the case $(A_{-})$ just from
  the relation $C = (\vec{\lambda} \vec{C}_2)^2 - K (\vec{C}_2)^2 < 0$.
  This means
  that the matrix $(h_{ab})$ is not positive-definite, i.e. we
  have a certain number ($k \geq 1$) of ``phantom'' fields among $\varphi^a$.

  Let us  consider the case $K > 0$, or equivalently,
  $\vec{\lambda}^2 > \lambda^2_0$. We get from (\ref{4.jd})
  $C > 0$.

   The solution
   is a non-singular one for $\tau \in ( -  \infty, + \infty )$
   in such two cases
   \bear{6.8}
      (B_{+}) \quad | \lambda_a C_2^a | <  \sqrt{C};
              \\ \label{6.9}
      (B_{0}) \quad | \lambda_a C_2^a |  = \sqrt{C}.
    \ear

   Indeed, for $K > 0$ the solution is defined for
   $u \in  ( -  \infty, + \infty )$ and the function
   (\ref{6.2}) has the following asymptotical behavior
    \beq{6.10b}
    \bar{f} \sim \mp (u -u_0)\sqrt{C} + \lambda_a C_2^a u + {\rm const}
    \eeq
    for $u \to \pm \infty$. This relation implies the divergency
    of the integral (\ref{6.3}) for $u = \pm \infty $, when
    restrictions $(B_{+})$ and $(B_{0})$ are imposed.

    In the case $(B_{+})$
    we get a bouncing behavior of  scale factor $a(\tau)$
    analogous to that in the case $(A_{+})$.
    In the case $(B_{0})$  the solution is
    non-singular, the function  $a(\tau)$ is monotonically
    increasing to infinity from a some non-zero  value
    (as in the case $(A_{0})$).

    For the $(B_{+})$-case we get
         \beq{6.11}
           (\lambda_a C_2^a)^2  <  C =(\vec{\lambda} \vec{C}_2)^2 - K
           (\vec{C}_2)^2,
     \eeq
  and, hence, $(\vec{C}_2)^2 = h_{ab} C_2^a C_2^b < 0$.
  The matrix $(h_{ab})$ is not positive-definite.

 Thus, we have  shown that the solutions with the bouncing behavior
 may take place only when there are ``phantom'' fields among the scalar ones.

 For non-singular solutions of the cases $(A_{0})$ and $(B_{0})$
 we get by analogous consideration that $(\vec{C}_2)^2 = h_{ab} C_2^a C_2^b =
 0$. In these cases the  non-singular solutions may take place also for
 positive-definite matrix  $(h_{ab})$ and but for zero constants: $C^a_2 =0$.

 %%%%%%%%%%%%%%%%%%%%%%%%%%%%%%%%%%%%%%%%%%%%%%%%%%%%%%%%%%%%%%%%%%%%%%
    \section{Conclusions}
 %%%%%%%%%%%%%%%%%%%%%%%%%%%%%%%%%%%%%%%%%%%%%%%%%%%%%%%%%%%%%%%%%%%%%%

 In this article we have studied a $(n+1)$-dimensional
 gravitational model with a set of
 $l$ scalar fields and an antisymmetric  form field of rank $p+2$.
 The special case of this model containing one scalar field ($l =1$)
 with positive-definite kinetic term was investigated in
 \cite{IS}.

 The metric was taken to be diagonal one, and for the $(p+2)$-form
 a composite electric $Sp$-brane form was adopted. All ansatz functions
 depended only on the one distinguished coordinate, $u$. Under these conditions the
  initial model was reduced to an effective 1-dimensional
 $\sigma$-model.  The diagonal form of our metric ansatz resulted
 in the  appearance  of constraint equations on the  charge
 densities of branes associated with the $(p+2)$-form field.

 It was found in \cite{IS} that
 these constraint equations may be satisfied for certain odd values of the
 space-time dimension, given by $D=4m+1 =5, 9, 13...$, the model
 allowed the maximal number of the charged electric branes.

 For these special odd dimensions and
 non-exceptional dilatonic coupling
 ($\vec{\lambda}^2 \neq  \frac{n}{4(n-1)}$)
 we have found exact cosmological-type  solutions given in equations
 (\ref{4.pa})-(\ref{4.pc}).
 We have singled out a special solution given by (\ref{4.jc}).
 describing an attractor  power-law expansion
 ( see, (\ref{5.pa}), (\ref{5.pa}) and (\ref{5.pc})), e.g.
 with accelerated expansion of the scale factor.

 We have also found an asymptotical Kasner type behavior
 of the solutions for small ($\tau \to + 0$) and large ($\tau \to +
 \infty$) values of proper time when the matrix $(h_{ab})$ is
 positive-definite. In (\ref{5.sl}) we have presented the
 relation between  Kasner parameters (for scalar fields)
 in both asymptotics.

 Here  we also have found  non-singular
 solutions, e.g.   with bouncing behavior of the scale factors.
 We have shown that solutions with bouncing behaviour
  take place when there are  ``phantom''  scalar fields in the model.

 \begin{center}
 {\bf Acknowledgments}
 \end{center}

  The work of V.D.I. and V.N.M. was supported in part by a DFG grant
  Nr. 436 RUS 113/807/0-1 and by the Russian Foundation for
   Basic Researchs, grant Nr. 05-02-17478.

  V.D.I. thanks colleagues from the Physical Department of
  the University of Konstanz for their hospitality during his visit
  in May-July and V.N.M. - during his visit in April-June, 2005.

   %\np

  \end{document}